\documentclass[aps,twocolumn,superscriptaddress,floatfix,prl]{revtex4}
\usepackage{graphicx,subfigure,amsmath,bbm}
\begin{document}
\title{Nuclear Spin Effect in Metallic Spin Valve}
\date{\today}

\author{J. Danon}
\author{Yu.\ V. Nazarov}
\affiliation{Kavli Institute of NanoScience, Delft University of Technology, 2628 CJ Delft, The Netherlands}
\begin{abstract}
We study electronic transport through a ferromagnet normal-metal ferromagnet system and we investigate the effect of hyperfine interaction between electrons and nuclei in the normal-metal part. A switching of the magnetization directions of the ferromagnets causes nuclear spins to precess. We show that the effect of this precession on the current through the system is large enough to be observed in experiment.
\end{abstract}
\maketitle


In recent years considerable theoretical and experimental work is aimed at the controlled manipulation of electron spin in nanoscale solid state systems, a field commonly referred to as \textit{spintronics}~\cite{dassarma}. The main motivations for this research are applications in conventional computer hardware~\cite{parkin} as well as the futuristic possibility of \textit{quantum computation}~\cite{awschalom}, using single electron spins as information carrying units (\textit{qubits}). For both purposes, understanding the mechanisms of spin polarization, relaxation, and dephasing in solid state systems is crucial.

The branch of metallic spintronics has quickly evolved after the discovery of the giant magnetoresistance (GMR) in hybrid ferromagnetic normal metal structures~\cite{baibich,binasch}. Theoretical and experimental studies on magnetic mutlilayers have not only revealed interesting physics, but also already led to several applications in magnetoelectronic devices. Magnetic recording read heads based on the GMR were first developed some ten years ago~\cite{gmrreadhead}, but nowadays can be found in nearly all hard disk drives.

In the context of quantum computation, semiconductor quantum dots are regarded as promising candidates for storing electron spin based qubits~\cite{lossdivincenzo}. Recent progress in quantum manipulation of single spins~\cite{petta:science} has overcome the effects of various spin relaxation processes in these devices. The unavoidable hyperfine interaction between electron and nuclear spin presently attracts much attention. It has been identified as the main source of spin relaxation in high-purity samples at low temperatures~\cite{merkulov:205309,erlingsson:195306} and can even govern the electron transport in double dots~\cite{frank:science}. At present, hyperfine interaction is seen as the main obstacle to demonstrate quantum computation with electron spins in solid state devices.

In many other fields, for instance nuclear magnetic resonance (NMR) experiments, hyperfine interactions play a central role already for decades. The Overhauser effect~\cite{overhauser:1953} is a common way to increase the degree of nuclear polarization in metals enhancing NMR peaks. In semiconductors, optical orientation techniques~\cite{opticalorientation} are used to polarize the nuclear system. In the context of \textit{metallic} devices, hyperfine interaction has been thought to be too weak to influence charge transport directly, and it has been regarded merely as an extra source of spin relaxation~\cite{dassarma}. 


In this paper, we predict a clearly observable hyperfine effect on electron transport in a \textit{metallic} device. Thereby we demonstrate that hyperfine interactions may be important and possibly even dominant also for metallic spintronic devices. We consider electronic transport through a ferromagnet normal-metal ferromagnet multilayer. This so-called \textit{spin valve} is the basic magnetoelectronic device and the core component of all GMR based read heads. By changing the magnetization directions of the two ferromagnetic leads, one alters the total resistance of the device as well as the degree and direction of electronic polarization in the normal metal part in the presence of a current.

Although the spin and particle transport properties of spin valves are well investigated and understood~\cite{spintransport}, effects of hyperfine interaction in magnetic multilayers have been hardly studied at all. One may think that these effects are negligible owing to the small value of the hyperfine interaction constant $A \simeq 10^{-6}$~eV in metals. We show, however, that electron spins accumulating in the normal metal part can build up a significant polarization of nuclear spins. The direction of this polarization is determined by the magnetizations of the leads. If the magnetizations are suddenly changed, this affects the electronic spin distribution in the normal metal part immediately (at a time scale $\tau_e \simeq 10^{-11}$ s). The nuclear spin polarization reacts on a much longer time scale and will start to precess slowly around its new equilibrium direction. In this paper we are mainly interested in the feedback of the nuclear polarization on the electronic system. We show that due to such feedback the precession manifests itself as oscillations in the net current through the device. The amplitude of these oscillations is estimated as $A/E_{th}$. Here $E_{th}$ is the Thouless energy characterizing the typical electron dwell time in the valve. The estimation is valid provided this time is shorter than the spin relaxation time $\tau_{sf}$, which sets an upper bound for the effect, $A\tau_{sf}/\hbar$.  For typical parameters, the relative magnitude of the current oscillations can be of the order $10^{-4}\sim 10^{-5}$, which is clearly large enough to be measured in experiments.


\begin{figure}[t]
\includegraphics{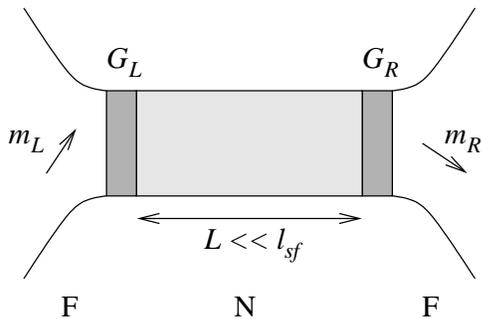}
\caption{A schematic picture of the system considered. A small metallic island is connected to two large ferromagnetic reservoirs with magnetizations $\vec m_L$ and $\vec m_R$. The contacts are charcterized by conductances $G_L$ and $G_R$, which consist of spin-dependent parts $G^\uparrow$ and $G^\downarrow$, and a mixing conductance $G^{\uparrow\downarrow}$. The length of the normal metal part is significantly smaller than the spin diffusion length.}\label{fig:fnf}
\end{figure}
We model our system as a small metallic island connected to two ferromagnetic leads (Fig.~\ref{fig:fnf}). We assume the island to be smaller than the spin diffusion length $l_{sf}$ and the time an electron spends in the island much smaller than $\tau_{sf}$, which allows us to disregard spin-orbit relaxation mechanisms in the island. We also assume that the resistance of the junctions by far exceeds the resistance of the island itself. In this case, we can describe the electronic states in the island with a single coordinate-independent distribution function $f(E,t)$.

The two ferromagnetic leads are modeled as large reservoirs in local equilibrium, with magnetizations in arbitrary directions $\vec m_L$ and $\vec m_R$. Assuming for simplicity $T=0$, we approximate the electronic distribution function in the leads as $f_{L(R)}(E)=\theta(\mu_{L(R)}-E)$. The difference in chemical potentials is due to the bias voltage applied $eV_b=\mu_L-\mu_R$. We can disregard temperature provided $eV_b\gg k_BT$.

In our model, the electron spin polarization  is mainly determined by the balance of spin-polarized currents flowing into and out of the ferromagnetic leads. However, a significant correction to this balance comes from  hyperfine coupling between the electron and nuclear spins. The resulting change of the polarization affects the net electric current in the device. So we will first derive an expression for hyperfine induced polarization of electrons and nuclei, and then we combine the result with the known expressions for spin transport through spin valves.

The Hamiltonian we use to describe the electronic and nuclear states in the island consists of an electronic part and a part describing the hyperfine interactions,
\begin{equation}\label{totham}
\begin{split}
\hat{H} &= \hat{H}_{el} +  \hat{H}_{hf} \\
\hat{H}_{el} &= \sum_k \epsilon_k \left( \hat{a}_{k\uparrow}^\dagger\hat{a}_{k\uparrow}+ \hat{a}_{k\downarrow}^\dagger\hat{a}_{k\downarrow}\right) \\
\hat{H}_{hf} &= \sum_n A_n \hat{\vec{S}}_n(t) \cdot \hat{\Psi}^\dagger(\vec r_n)\;\frac{1}{2}\vec{\sigma}\;\hat{\Psi}(\vec r_n),
\end{split}
\end{equation}
where $\hat{a}_{k\alpha}^{(\dagger)}$ are electron annihilation (creation) operators for spin up and down ($\alpha=\uparrow,\downarrow$). We expressed the usual hyperfine contact Hamiltonian in electronic field operators, defined as $\hat{\Psi}(\vec r)=\Omega^{-1/2}\sum_{k,\alpha}\hat{a}_{k\alpha}e^{ik\vec r}$, where $\Omega$ is the volume of the island. $A_n$ is the hyperfine coupling coefficient between an electron and the nucleus at position $\vec r_n$, the vectors $\hat{\vec S}_n$ are the nuclear spin operators and $\vec\sigma$ the Pauli spin matrices.


We apply a second order perturbation expansion to find an expression for the time dependence of the electronic distribution function $f(k,t)=\langle \hat{a}_{k\alpha}^\dagger(t)\hat{a}_{k\beta}(t)\rangle$
\begin{multline}\label{exp}
\frac{df(k,t)}{dt}\approx \left\langle \frac{i}{\hbar}\left[ \hat{H}(t),\hat{a}_{k\alpha}^\dagger\hat{a}_{k\beta}\right]\right\rangle\\
-\left\langle \int_{-\infty}^t \frac{dt'}{\hbar^2} \left[ \hat{H}(t),\left[  \hat{H}(t'),\hat{a}_{k\alpha}^\dagger\hat{a}_{k\beta}\right] \right]\right\rangle,
\end{multline}
where the indices $\alpha$ and $\beta$ now span a $2\times 2$ spin space. We see that the expansion can be completely expressed in the commuting operators $\hat{\vec{S}}_n$ and $\hat{a}^{(\dagger)}$.

Using Wick's theorem, we write the terms with four and six creation and annihilation operators as products of pairs, which then again can be interpreted as distribution functions $f(E,t)$. Further, we assume that the electrons are distributed homogeneously on the island and approximate $ A_n = A/n_0$, $A$ being the hyperfine coupling energy of the material and $n_0$ the density of nuclei with non-zero spin~\cite{hyperfinerev}. We find up to the second order
\begin{equation}\label{dfdt}
\begin{split}
\left( \frac{d\vec f_s}{dt}\right)_{hf} = &\phantom{-} \frac{A}{\hbar}\langle\vec{S}(t)\rangle\times\vec f_s(t) \\
&- \frac{1}{\tau_{hf}} \left[ \frac{1}{2}-\vec f_s(t)\cdot\langle\vec{S}(t)\rangle\right]\vec f_s(t) \\
&+ \frac{1}{\tau_{hf}} f_p(t)\left[ 1-f_p(t)\right] \langle\vec{S}(t)\rangle,
\end{split}
\end{equation}
where we used the notation $f=f_p\mathbbm{1}+\vec f_s\cdot \vec \sigma$, i.e.\ we split $f$ in a particle and spin part. Nuclear spin enters here as the average polarization $\langle\vec{S}(t)\rangle$. The first-order term describes the precession of electron spin in the field of the nuclei and disappears if electron and nuclear polarizations are aligned. The second-order terms are all proportional to the hyperfine relaxation time defined as $\tau_{hf}=\hbar n_0/\pi A^2\nu$, $\nu$ being the density of states at the Fermi energy. Since  hyperfine interaction affects the electron spin only, the contribution to the time derivative of $f_p$ is zero.

This contribution is not the main one in the balance in the spin valve. Mainly, it is determined by electron transfers through the spin-active junctions. To describe this, we use the approach of \cite{spintransport} that is valid for non-collinear magnetizations. This yields
\begin{equation}\label{defeq}
\begin{split}
\left( \frac{d\vec f_s}{dt}\right)_{sv}= &\; \vec B_{e}\times \vec f_s(t)-\frac{1}{\tau_e} \vec f_s(t)\\
&+ \frac{1}{\tau_e}
\Big\{ \alpha_L\left[ 1-f_p(t)\right] +\beta_L\vec m_L\cdot\vec f_s(t) \Big\}\vec m_L\\
&+ \frac{1}{\tau_e}
\Big\{ \alpha_R\left[ -f_p(t)\right] +\beta_R\vec m_R\cdot\vec f_s(t) \Big\}\vec m_R. 
\end{split}
\end{equation}
Following \cite{spintransport}, we describe each spin-active junction with four conductances, $G^{\uparrow}$, $G^{\downarrow}$ and $G^{\uparrow\downarrow} = (G^{\downarrow\uparrow})^{*}$. (If the junction is not spin-active, $G^{\uparrow} = G^{\downarrow} = G^{\uparrow\downarrow} = G^{\downarrow\uparrow} = G$). The electron spin is subject to an effective field
\begin{equation*}
\vec B_{e} =-\frac{1}{\tau_e}\frac{\mathrm{Im}(G^{\uparrow\downarrow}_L)\vec m_L + \mathrm{Im}(G^{\uparrow\downarrow}_R)\vec m_R}{\mathrm{Re}(G^{\uparrow\downarrow}_L + G^{\uparrow\downarrow}_R)},
\end{equation*}
and we introduce dimensionless parameters characterizing the spin activity of the junctions
\begin{equation*}
\begin{split}
\alpha_{L(R)} &=\frac{G^{\uparrow}_{L(R)}-G^{\downarrow}_{L(R)}}{2\mathrm{Re}(G^{\uparrow\downarrow}_L + G^{\uparrow\downarrow}_R)},\quad P_{L(R)} =\frac{G^{\uparrow}_{L(R)}-G^{\downarrow}_{L(R)}}{G^{\uparrow}_{L(R)}+G^{\downarrow}_{L(R)}}\\
\beta_{L(R)}&=\frac{2\mathrm{Re}(G_{L(R)}^{\uparrow\downarrow})-G^{\uparrow}_{L(R)}-G^{\downarrow}_{L(R)}}{2\mathrm{Re}(G^{\uparrow\downarrow}_L + G^{\uparrow\downarrow}_R)}.
\end{split}
\end{equation*}
The order of magnitude of (\ref{defeq}) is estimated as $1/\tau_e$, $\tau_e$ being a typical electron dwell time in the island, $\tau_e = e^2\nu \Omega/2\mathrm{Re}(G^{\uparrow\downarrow}_L + G^{\uparrow\downarrow}_R)$. 

Since by the essence of the spin valve, the spin balance affects the particle current through the device, the time derivative of $f_p$ is non-zero now,
\begin{equation}\label{df0dt}
\begin{split}
\frac{df_p}{dt} = &\frac{1}{\tau_e}\big\{ \frac{\alpha_L}{P_L} \left[1-f_p(t)\right]-\frac{\alpha_R}{P_R} f_p(t)\\
&\quad-(\alpha_L \vec m_L+\alpha_R\vec m_R)\cdot\vec f_s(t)\big\}.
\end{split}
\end{equation}

We also need an equation for the time dependence of the nuclear spin polarization. One can obtain it from a perturbation expansion similar to 
(\ref{exp}) or directly inherit it from the fact that
hyperfine interaction conserves the total spin of electrons and
nuclei
\begin{equation}\label{dsdt}
n_0\left(\frac{d\langle\vec{S}\rangle}{dt}\right)_{hf}= -\nu eV_b\left( \frac{d\vec f_s}{dt}\right)_{hf}.
\end{equation}
We see from this that the relaxation time of nuclear spins  $\tau_d \simeq \tau_{hf} n_0/\nu eV_b$. We assume that no other relaxation mechanism provides a shorther relaxation time.

Let us now estimate and compare the time scales involved. For the nuclear system, the precession frequency $\omega = A\nu eV_b\vert\vec f_s\vert/\hbar n_0$ and the relaxation time  $\tau_d\simeq \hbar n_0^2/\pi A^2\nu^2eV_b$. Typical values for $A$ range from $10^{-7}\:\mathrm{eV}$ for weak coupling to $10^{-4}\:\mathrm{eV}$ (e.g.\ in GaAs~\cite{hyperfinerev}); we chose $A= 5\times 10^{-6}\:\mathrm{eV}$.  We take typical solid-state parameters to estimate $n_0= 2.9\times 10^{29}$~m$^{-3}$ and $\nu= 1.3\times 10^{47}$~J$^{-1}$m$^{-3}$.  For an applied bias voltage of $V_b=10\:\mathrm{mV}$, this results in a precession frequency $\omega\sim 10^5\:\mathrm{Hz}$ and a nuclear relaxation time $\tau_d\sim 0.5\:\mathrm{s}$. For the electronic system, the spin relaxation rate consists of two terms $\propto$ $1/\tau_e$ and $\propto 1/\tau_{hf}$. We set the conductance of the F/N-interfaces to $G\sim 3\ \Omega^{-1}$~\cite{xia:220401}. If we choose dimensions of the metal island of $0.1\times 0.1\times 5\:\mu\mathrm{m}^3$, we find $\tau_e\sim 10^{-11}$~s. This corresponds to a Thouless energy $E_{th} = \hbar/\tau_e \simeq 0.06$~meV. The estimation for the hyperfine relaxation time reads $\tau_{hf}\sim 10^{-4}$~s.

We conclude that on the time scale of all nuclear processes, $\vec f_s$ and $f_p$ instantly adjust themselves to current values of voltage, magnetization, and importantly, nuclear spin polarization. Their values are determined from the spin balance,
\begin{equation}
\left( \frac{d\vec f_s}{dt}\right)_{hf}+\left( \frac{d\vec f_s}{dt}\right)_{sv}=0\quad\text{and}\quad\frac{df_p}{dt} = 0.\label{fpeq}
\end{equation}
As to nuclear polarization at constant voltage and magetization, it is of the order of $1$ owing to a sort of Overhauser effect produced  by non-equilibrium electrons passing the island. Indeed, it follows from Eq.\ \ref{dsdt} that the stationary $2 \langle\vec S\rangle = \vec f_s /[f_p(1-f_p) + \vert\vec f_s\vert^2]$. We see that $\langle\vec S\rangle$ and $\vec f_s$ are parallel under stationary conditions. This is disappointing since this will not result in any precession.

The essential ingredient of our proposal is to change in time the magnetization(s) of the leads. Let us consider the effect of sudden change of the magnetization in one of the leads at $t=0$. The electrons will find their new distribution, characterized by $\vec f_{new}$, on a timescale of $\tau_e$. As we see from (\ref{dsdt}), the nuclear spin system will start to precess around $\vec f_{new}$ with the frequency estimated. The precessing polarization will contribute to the effective field $\vec B_e$, $\vec B_e \to \vec B_e + A\langle\vec S(t)\rangle/\hbar$ in (\ref{fpeq}). This will result in a small correction to $\vec f_{new}$, $\vec f^{(1)}$, which is visible in the net current through the junction, due to its oscillating nature. A simple expression for this correction is obtained in the limit of weakly polarizing junctions ($\alpha_{L,R},\beta_{L,R} \ll 1$),
\begin{equation}\label{ffo}
\vec f^{(1)}(t)=\frac{A\tau_e}{\hbar}\langle\vec S(t)\rangle\times\vec f_{new} \quad\propto\quad A/E_{th},
\end{equation}
while a more general expression is obtained by solving (\ref{fpeq}) up to first order in $A$. The oscillatory part of the resulting current is given by
\begin{equation}\label{current}
\tilde I(t) = V_b\frac{e^2\nu\Omega}{2\tau_e}[P_R\vec m_R-P_L\vec m_L]\cdot\vec f^{(1)}(t).
\end{equation}
The time dependence of nuclear polarization is still governed by Eq. \ref{dsdt}.

Combining Eqs \ref{ffo} and \ref{current}, we find that the time-dependent current follows the behavior of $\langle\vec S(t)\rangle$ and therefore exhibits oscillations with frequency $\omega$ that are damped at the long time scale $\tau_d$. The amplitude of these oscillations $\Delta I$ in the limit of small $\alpha$ and $\beta$ reads
\begin{equation}\label{amp}
\frac{\vert\Delta I\vert}{\langle I\rangle}\approx
\frac{A\tau_e}{\hbar}\vert\langle\vec S_{old}\rangle\times\vec f_{new}\vert\vert P_R\vec m_R^\perp-P_L\vec m_L^\perp\vert ,
\end{equation}
again proportional to $A/E_{th}$. In this equation $\langle\vec S_{old}\rangle$ refers to the nuclear polarization before switching the magnetizations and $\vec m_{L(R)}^\perp$ is the part of $\vec m_{L(R)}$ perpendicular to  $\vec f_{new}$. This relation makes it straightforward that one needs non-collinear  magnetizations to observe any effect.

In the same limit, the damping time and precession frequency are given by
\begin{equation}\label{rel}
\tau_d =\tau_{hf} \frac{n_0}{\nu eV_b}\left\lbrace\xi+\xi^{-1}+2\right\rbrace,
\end{equation}
and
\begin{equation}\label{pre}
\omega =\frac{A}{\hbar}\frac{\nu eV_b}{n_0}\frac{\vert P_R\vec m_R^\perp-P_L\vec m_L^\perp\vert}{\xi+\xi^{-1}+2}.
\end{equation}
where $\xi = G_L/G_R$ characterizes the asymmetry of the conductance of the contacts.

\begin{figure}[t]
\includegraphics[width=8.5cm]{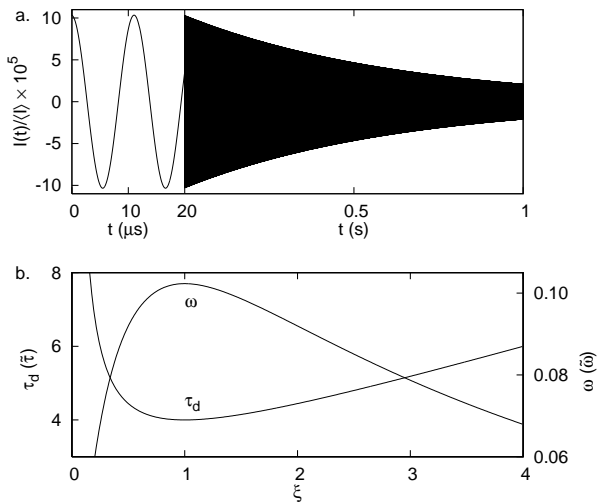}
\caption{(a) Numerical calculation of the relative current fluctuations as a function of time. For $t>20$~$\mu$s the $t$-axis is compressed. We used the parameters chosen in the text and further we took for both contacts $\alpha=0.128$, $\beta=0.115$ and $P=0.333$. The magnetizations switch at $t=0$ from $m_L=(0,0,1)$, $m_R=(0,0,-1)$ to $m_L=(0,1,0)$, $m_R=(0,0,-1)$. (b) The dependence of the relaxation time and the frequency of the fluctuations on the asymmetry $\xi$ in the conductances, where $\tilde \tau=\hbar n_0/\pi A^2\nu^2eV_b$ and $\tilde\omega=A\nu eV_b/\hbar n_0$.}\label{fig:plot1}
\end{figure}
In Fig.\ \ref{fig:plot1}a we plotted a numerical solution for the current $\tilde I(t)/\langle I\rangle$ and in \ref{fig:plot1}b the dependence of $\omega$ and $\tau_d$ on the asymmetry in conductance of the contacts. For \ref{fig:plot1}a we made use of equations (\ref{dsdt}) and (\ref{fpeq}), and inserted realistic $\alpha_{L,R}$ and $\beta_{L,R}$. For the parameters used, the estimate (\ref{amp}) of the amplitude is 4.4$\times10^{-5}$. Eqs\ (\ref{rel}) and (\ref{pre}) give $\tau_d=0.63$~s and $\omega=1.0\times 10^5\:\mathrm{Hz}$, in agreement with the plot. Typical currents through spin valves of these dimensions using a bias voltage of 10~mV range between 10 and 100~mA. Oscillations of the order of $10^{-5}$ - $10^{-4}$ should be clearly visible in experiment. The unavoidable shot noise due to the discrete nature of the electrons crossing the junctions will not prevent even an accurate single-shot measurement, since the measurement time can be of the order of $\tau_d$. An estimate using $(\delta I)^2\simeq 2eI/\tau_d$ gives a relative error of $10^{-8}$ - $10^{-9}$, at least three orders smaller than the oscillations. 

So far we have assumed precisely uniform electron distributions. In a realistic situation however, the finite resistance of the island results in a voltage drop over the island, thus causing spatial variation of $f_0$ and $\vec f_s$. Importantly, this gives variations in the precession frequency $\omega\propto A\vert\vec f_s\vert/\hbar$. Such variation $\Delta\omega$ over the length of the island will contribute to an apparent relaxation of the spin polarization, since precession in different points of the island occurs with a slightly different frequency. This effect adds a term $1/\tau^\ast=\Delta\omega$ to the damping rate $1/\tau_d$. Assuming a simple linear voltage drop over the normal metal part, we find $1/\tau^\ast=(G_{\mathrm{junc}}/G_{\mathrm{isl}})\omega_0$, i.e.\ the ratio of the total conductance of the spin valve and the conductance of the metal island times the average oscillation frequency $\omega_0$. Although the effect can reduce the apparent relaxation time, provided $(\Delta \omega) \tau_d \gg 1$, it will not influence the time-dependent current just after $t=0$.


In conclusion, we have shown how hyperfine-induced nuclear precession in the normal metal part of a spin valve can be made experimentally visible. The precession should give a clear signature in the form of small oscillations in the net current through the valve after sudden change of the magnetizations of the leads. We found a coupled set of equations describing the nuclear and electron spin dynamics resulting from a second order perturbation expansion in the hyperfine contact Hamiltonian. We presented a numerical solution for the net current and derived an estimate for the amplitude of the oscillations. We found that the relative amplitude of these oscillations is sufficiently big to be observable.

The authors acknowledge financial support from FOM and useful discussions with G.E.W.\ Bauer.

\bibliography{paper}
\end{document}